\documentclass[%
 aip,
 jmp,%
 amsmath,amssymb,
 superscriptaddress,
 reprint,%
]{revtex4}

\usepackage{amsmath,amsfonts,amssymb,color,epsfig,graphics,graphicx,latexsym,revsymb,theorem,url,verbatim,subfigure}

\newtheorem{definition}{Definition}
\newtheorem{proposition}[definition]{Proposition}
\newtheorem{lemma}[definition]{Lemma}

\newtheorem{theorem}[definition]{Theorem}
\newtheorem{corollary}[definition]{Corollary}
\newtheorem{conjecture}[definition]{Conjecture}

\newtheorem{remark}[definition]{Remark}
\newtheorem{example}[definition]{Example}
\newtheorem{question}[definition]{Question}

\newenvironment{proof}{\noindent \textbf{{Proof.~} }}{\qed}
\def\Dbar{\leavevmode\lower.6ex\hbox to 0pt
{\hskip-.23ex\accent"16\hss}D}
\makeatletter
\def\url@leostyle{%
  \@ifundefined{selectfont}{\def\UrlFont{\sf}}{\def\UrlFont{\small\ttfamily}}}
\makeatother
\def\bcj{\begin{conjecture}}
\def\ecj{\end{conjecture}}
\def\bcr{\begin{corollary}}
\def\ecr{\end{corollary}}
\def\bd{\begin{definition}}
\def\ed{\end{definition}}
\def\bea{\begin{eqnarray}}
\def\eea{\end{eqnarray}}
\def\bem{\begin{enumerate}}
\def\eem{\end{enumerate}}
\def\bex{\begin{example}}
\def\eex{\end{example}}
\def\bim{\begin{itemize}}
\def\eim{\end{itemize}}
\def\bl{\begin{lemma}}
\def\el{\end{lemma}}
\def\bpf{\begin{proof}}
\def\epf{\end{proof}}
\def\bpp{\begin{proposition}}
\def\epp{\end{proposition}}
\def\bqu{\begin{question}}
\def\equ{\end{question}}
\def\br{\begin{remark}}
\def\er{\end{remark}}
\def\bt{\begin{theorem}}
\def\et{\end{theorem}}

\def\btb{\begin{tabular}}
\def\etb{\end{tabular}}

\newcommand{\nc}{\newcommand}

 \nc{\bA}{{\bf A}} \nc{\bB}{{\bf B}} \nc{\bC}{{\bf C}}
 \nc{\bD}{{\bf D}} \nc{\bE}{{\bf E}} \nc{\bF}{{\bf F}}
 \nc{\bG}{{\bf G}} \nc{\bH}{{\bf H}} \nc{\bI}{{\bf I}}
 \nc{\bJ}{{\bf J}} \nc{\bK}{{\bf K}} \nc{\bL}{{\bf L}}
 \nc{\bM}{{\bf M}} \nc{\bN}{{\bf N}} \nc{\bO}{{\bf O}}
 \nc{\bP}{{\bf P}} \nc{\bQ}{{\bf Q}} \nc{\bR}{{\bf R}}
 \nc{\bS}{{\bf S}} \nc{\bT}{{\bf T}} \nc{\bU}{{\bf U}}
 \nc{\bV}{{\bf V}} \nc{\bW}{{\bf W}} \nc{\bX}{{\bf X}}
 \nc{\bZ}{{\bf Z}}

\nc{\cA}{{\cal A}} \nc{\cB}{{\cal B}} \nc{\cC}{{\cal C}}
\nc{\cD}{{\cal D}} \nc{\cE}{{\cal E}} \nc{\cF}{{\cal F}}
\nc{\cG}{{\cal G}} \nc{\cH}{{\cal H}} \nc{\cI}{{\cal I}}
\nc{\cJ}{{\cal J}} \nc{\cK}{{\cal K}} \nc{\cL}{{\cal L}}
\nc{\cM}{{\cal M}} \nc{\cN}{{\cal N}} \nc{\cO}{{\cal O}}
\nc{\cP}{{\cal P}} \nc{\cQ}{{\cal Q}} \nc{\cR}{{\cal R}}
\nc{\cS}{{\cal S}} \nc{\cT}{{\cal T}} \nc{\cU}{{\cal U}}
\nc{\cV}{{\cal V}} \nc{\cW}{{\cal W}} \nc{\cX}{{\cal X}}
\nc{\cZ}{{\cal Z}}

\nc{\hA}{{\hat{A}}} \nc{\hB}{{\hat{B}}} \nc{\hC}{{\hat{C}}}
\nc{\hD}{{\hat{D}}} \nc{\hE}{{\hat{E}}} \nc{\hF}{{\hat{F}}}
\nc{\hG}{{\hat{G}}} \nc{\hH}{{\hat{H}}} \nc{\hI}{{\hat{I}}}
\nc{\hJ}{{\hat{J}}} \nc{\hK}{{\hat{K}}} \nc{\hL}{{\hat{L}}}
\nc{\hM}{{\hat{M}}} \nc{\hN}{{\hat{N}}} \nc{\hO}{{\hat{O}}}
\nc{\hP}{{\hat{P}}} \nc{\hR}{{\hat{R}}} \nc{\hS}{{\hat{S}}}
\nc{\hT}{{\hat{T}}} \nc{\hU}{{\hat{U}}} \nc{\hV}{{\hat{V}}}
\nc{\hW}{{\hat{W}}} \nc{\hX}{{\hat{X}}} \nc{\hZ}{{\hat{Z}}}

\newcommand{\bra}[1]{\langle#1|}
\newcommand{\ket}[1]{|#1\rangle}

\def\Dbar{\leavevmode\lower.6ex\hbox to 0pt
{\hskip-.23ex\accent"16\hss}D}
\begin{document}


\def\be{\begin{eqnarray}}
\def\ee{\end{eqnarray}}


\newcommand{\ca}{\mathcal A}

\newcommand{\cb}{\mathcal B}
\newcommand{\cc}{\mathcal C}
\newcommand{\cd}{\mathcal D}
\newcommand{\ce}{\mathcal E}
\newcommand{\cf}{\mathcal F}
\newcommand{\cg}{\mathcal G}
\newcommand{\ch}{\mathcal H}
\newcommand{\ci}{\mathcal I}
\newcommand{\cj}{\mathcal J}
\newcommand{\ck}{\mathcal K}
\newcommand{\cl}{\mathcal L}
\newcommand{\cm}{\mathcal M}
\newcommand{\cn}{\mathcal N}
\newcommand{\co}{\mathcal O}
\newcommand{\cp}{\mathcal P}
\newcommand{\cq}{\mathcal Q}
\newcommand{\calr}{\mathcal R}
\newcommand{\cs}{\mathcal S}
\newcommand{\ct}{\mathcal T}
\newcommand{\cu}{\mathcal U}
\newcommand{\cv}{\mathcal V}
\newcommand{\cw}{\mathcal W}
\newcommand{\cx}{\mathcal X}
\newcommand{\cy}{\mathcal Y}
\newcommand{\cz}{\mathcal Z}


\newcommand{\sa}{\mathscr{A}}
\newcommand{\sm}{\mathscr{M}}


\newcommand{\fa}{\mathfrak{a}}  \newcommand{\Fa}{\mathfrak{A}}
\newcommand{\fb}{\mathfrak{b}}  \newcommand{\Fb}{\mathfrak{B}}
\newcommand{\fc}{\mathfrak{c}}  \newcommand{\Fc}{\mathfrak{C}}
\newcommand{\fd}{\mathfrak{d}}  \newcommand{\Fd}{\mathfrak{D}}
\newcommand{\fe}{\mathfrak{e}}  \newcommand{\Fe}{\mathfrak{E}}
\newcommand{\ff}{\mathfrak{f}}  \newcommand{\Ff}{\mathfrak{F}}
\newcommand{\fg}{\mathfrak{g}}  \newcommand{\Fg}{\mathfrak{G}}
\newcommand{\fh}{\mathfrak{h}}  \newcommand{\Fh}{\mathfrak{H}}
\newcommand{\fraki}{\mathfrak{i}}       \newcommand{\Fraki}{\mathfrak{I}}
\newcommand{\fj}{\mathfrak{j}}  \newcommand{\Fj}{\mathfrak{J}}
\newcommand{\fk}{\mathfrak{k}}  \newcommand{\Fk}{\mathfrak{K}}
\newcommand{\fl}{\mathfrak{l}}  \newcommand{\Fl}{\mathfrak{L}}
\newcommand{\fm}{\mathfrak{m}}  \newcommand{\Fm}{\mathfrak{M}}
\newcommand{\fn}{\mathfrak{n}}  \newcommand{\Fn}{\mathfrak{N}}
\newcommand{\fo}{\mathfrak{o}}  \newcommand{\Fo}{\mathfrak{O}}
\newcommand{\fp}{\mathfrak{p}}  \newcommand{\Fp}{\mathfrak{P}}
\newcommand{\fq}{\mathfrak{q}}  \newcommand{\Fq}{\mathfrak{Q}}
\newcommand{\fr}{\mathfrak{r}}  \newcommand{\Fr}{\mathfrak{R}}
\newcommand{\fs}{\mathfrak{s}}  \newcommand{\Fs}{\mathfrak{S}}
\newcommand{\ft}{\mathfrak{t}}  \newcommand{\Ft}{\mathfrak{T}}
\newcommand{\fu}{\mathfrak{u}}  \newcommand{\Fu}{\mathfrak{U}}
\newcommand{\fv}{\mathfrak{v}}  \newcommand{\Fv}{\mathfrak{V}}
\newcommand{\fw}{\mathfrak{w}}  \newcommand{\Fw}{\mathfrak{W}}
\newcommand{\fx}{\mathfrak{x}}  \newcommand{\Fx}{\mathfrak{X}}
\newcommand{\fy}{\mathfrak{y}}  \newcommand{\Fy}{\mathfrak{Y}}
\newcommand{\fz}{\mathfrak{z}}  \newcommand{\Fz}{\mathfrak{Z}}

\newcommand{\cfg}{\dot \fg}
\newcommand{\cFg}{\dot \Fg}
\newcommand{\ccg}{\dot \cg}
\newcommand{\circj}{\dot {\mathbf J}}
\newcommand{\circs}{\circledS}
\newcommand{\jmot}{\mathbf J^{-1}}


\newcommand{\rmd}{\mathrm d}
\newcommand{\mca}{\ ^-\!\!\ca}
\newcommand{\pca}{\ ^+\!\!\ca}
\newcommand{\peq}{^\Psi\!\!\!\!\!=}
\newcommand{\lt}{\left}
\newcommand{\rt}{\right}
\newcommand{\HN}{\hat{H}(N)}
\newcommand{\HM}{\hat{H}(M)}
\newcommand{\Hv}{\hat{H}_v}
\newcommand{\cyl}{\mathbf{Cyl}}
\newcommand{\lag}{\left\langle}
\newcommand{\rag}{\right\rangle}
\newcommand{\Ad}{\mathrm{Ad}}
\newcommand{\trace}{\mathrm{tr}}
\newcommand{\bbc}{\mathbb{C}}
\newcommand{\AC}{\overline{\mathcal{A}}^{\mathbb{C}}}
\newcommand{\Ar}{\mathbf{Ar}}
\newcommand{\uc}{\mathrm{U(1)}^3}
\newcommand{\M}{\hat{\mathbf{M}}}
\newcommand{\spin}{\text{Spin(4)}}
\newcommand{\id}{\mathrm{id}}
\newcommand{\Pol}{\mathrm{Pol}}
\newcommand{\Fun}{\mathrm{Fun}}
\newcommand{\bp}{p}
\newcommand{\act}{\rhd}
\newcommand{\data}{\lt(j_{ab},A,\bar{A},\xi_{ab},z_{ab}\rt)}
\newcommand{\datao}{\lt(j^{(0)}_{ab},A^{(0)},\bar{A}^{(0)},\xi_{ab}^{(0)},z_{ab}^{(0)}\rt)}
\newcommand{\deltadata}{\lt(j'_{ab}, A',\bar{A}',\xi_{ab}',z_{ab}'\rt)}
\newcommand{\background}{\lt(j_{ab}^{(0)},g_a^{(0)},\xi_{ab}^{(0)},z_{ab}^{(0)}\rt)}
\newcommand{\sgn}{\mathrm{sgn}}
\newcommand{\vth}{\vartheta}
\newcommand{\rmi}{\mathrm{i}}
\newcommand{\bfmu}{\pmb{\mu}}
\newcommand{\bfnu}{\pmb{\nu}}
\newcommand{\bfm}{\mathbf{m}}
\newcommand{\bfn}{\mathbf{n}}


\newcommand{\sz}{\mathscr{Z}}
\newcommand{\sk}{\mathscr{K}}


\title[Random Invariant Tensors]{Random Invariant Tensors}

\author{Youning Li}\thanks{Author to whom correspondence should be addressed: liyouning@mail.tsinghua.edu.cn}%
\affiliation{Department of Physics, Tsinghua University, Beijing, People's Republic of China}%
\affiliation{Collaborative Innovation Center of Quantum Matter, Beijing 100190, People's Republic of China}%
\author{Muxin Han}%
\affiliation{Department of Physics, Florida Atlantic University, FL 33431, USA}%
\affiliation{Institut f\"ur Quantengravitation, Universit\"at Erlangen-N\"urnberg, Staudtstr. 7/B2, 91058 Erlangen, Germany}

\author{Dong Ruan}
\affiliation{Department of Physics, Tsinghua University, Beijing, People's Republic of China}%
\affiliation{Collaborative Innovation Center of Quantum Matter, Beijing 100190, People's Republic of China}%

\author{Bei Zeng}
\affiliation{Department of Mathematics \& Statistics, University of
  Guelph, Guelph, Ontario, Canada}%
\affiliation{Institute for Quantum Computing and Department of Physics and Astronomy, University of Waterloo, Waterloo, Ontario, Canada}
\affiliation{Canadian Institute for Advanced Research, Toronto,
  Ontario, Canada}

\date{\today}

\begin{abstract}
Invariant tensors are states in the (local) SU(2) tensor product representation but invariant under global SU(2) action. They are of importance in the study of loop quantum gravity.
A random tensor is an ensemble of tensor states. An average over the ensemble is carried out when computing any physical quantities. The random tensor exhibits a phenomenon of `concentration of measure', saying that for any bipartition, the expected value of entanglement
entropy of its reduced density matrix is asymptotically the maximal possible as the local dimension goes to infinity. This is also true even when the average is over the invariant subspace instead of the whole space for $4-$valent tensors, although its entropy deficit is divergent. One might expect that for $n\geq 5$, $n-$valent random invariant tensor would behavior similarly. However, we show that, the expected entropy deficit of reduced density matrix of such $n-$valent random invariant tensor from maximum, is not divergent but a finite number. Under some special situation, the number could be even smaller than half a bit, which is the deficit of random pure state over the whole Hilbert space from maximum.
\end{abstract}

\maketitle
\renewcommand\theequation{\arabic{section}.\arabic{equation}}
\setcounter{tocdepth}{4}
\makeatletter
\@addtoreset{equation}{section}
\makeatother

\section{\label{sec:intro}Introduction}

An SU(2) $n$-valent invariant tensor $\psi$ is a state in the tensor product of SU(2) irreducible representation, and is invariant under the action of SU(2) group. The invariant tensor $\psi$ satisfies a quantum constraint equation $\sum_{i=1}^n\mathbf{\hat{J}}_i\psi=0$, where $\mathbf{\hat{J}}_i$ is the vector angular momentum operator $(\hat{J_x}, \hat{J_y}, \hat{J_z})$. The three operators $\hat{J_x}$, $\hat{J_y}$ and $\hat{J_z}$ constitute the su(2) Lie algebra generators acting at the $i$-th tensor component respectively. The invariant tensors play a central role in the theory of Loop Quantum Gravity (LQG) \cite{book,review,review1,rovelli2014covariant}, and particularly the structure of Spin-Networks
\cite{Penrose,Rovelli:1995ac,Baez:1994hx}. The spin-network state, as a quantum state of gravity, represents the quantization
of geometry in LQG. Classically, polyhedral geometries can be used for discretization of an arbitrary three-dimensional geometry. The geometries are quantized by the spin-network states at the quantum level \cite{Rovelli1995,ALvolume,ALarea}. As the
building block of spin-network, an SU(2) $n$-valent invariant tensor represents the quantum geometry of a polyhedron with $n$ faces. The three-dimensional quantum geometry is constructed by collecting a large number of invariant tensors representing different quantum geometrical polyhedra \cite{shape,CF,Li2016Invariant}. It corresponds to the kinematics of four-dimensional quantum gravity.

On the other hand, a \emph{random tensor} is an ensemble of tensor states, in which a random average is carried out in computing any physical quantities. In this paper we focus on the random tensor in Haar ensemble, i.e. the random average is an integral over all unitary transformations acting on the tensor with the Haar measure on the unitary group. The random tensors have recently attracted attention from quantum information theory, condensed matter theory, and quantum gravity \cite{Page1993Average,Roberts:2016hpo,Qi1,QiYY,Han:2017uco}. The random tensors are employed to construct the {Tensor Networks} to approximate Conformal Field Theory (CFT) states, which provide models to realize the AdS/CFT correspondence.




It has been well-known that, when the local system dimension goes to infinity, the subsystem entanglement entropy of a random tensor would approach very close to be maximum, the difference would be only one-half unit of information less \cite{hayden2006aspects}
\begin{eqnarray}
\overline{S} \rightarrow n\ln d-\frac{1}{2},\quad d\rightarrow \infty,
\end{eqnarray}
where $n$ is the number of particles that are not traced and $d$ is the dimension of the local subsystem.
In other words, a random tensor is close to a \emph{perfect tensor} when the local system dimension is large. A perfect tensor is a maximal entangled state for any bipartition of its local systems \cite{Pastawski:2015qua}. The concept of invariant perfect tensor (IPT) has been introduced \cite{Li2016Invariant} to understand the overlap of the concepts between invariant and perfect tensors. The nontrivial overlaps are found for tensors with rank 2 and 3. The existence of such IPT was finally proved to be false when the number of subsystem is greater than $4$ \cite{Chen2017Local}. Although the $4-$valent IPT does not exist, it is shown that a random rank-$4$ invariant tensor behaves approximately as a perfect tensor \cite{Li2016Invariant}
\begin{eqnarray}
\overline{S_2}&=&\ln \lt[(2j+1)^2+(2j+1)\rt]-\ln\lt(2\sum_{I=0}^{2j}(2I+1)^{-1}\rt),\notag\\
\overline{S_2}&\rightarrow &2\ln(2j+1)- \ln\ln (2j+1)-2\ln2-\epsilon_0,\quad j\rightarrow \infty
\end{eqnarray}
where $2j+1$ is the dimension of local subsystem and $\epsilon_0$ is the Euler constant.

The leading contribution of the entanglement entropy approaches the maximum as the local system dimension is large, however unlike the usual random tensor, when we restrict the ensemble to be the $4-$valent tensors in the SU(2) invariant subspace, the entanglement entropy of the random invariant tensor has a divergent subleading term \footnote{The leading term scales linearly to the dimension, while the subleading term is logarithmic to the dimension.}.


In this paper, we generalize the analysis of the random invariant tensor to rank $n>4$, and show that when $n \geq 5$, the subsystem entanglement entropy of rank-$n$ random tensor in the SU(2) invariant subspace approaches to the maximum asymptotically when the local system dimension is large. Interestingly, the random invariant tensor with $n>4$ is closer to the perfect tensor than $n=4$, since the entropy deficit is a finite number, in contrast to the divergent one in the $n=4$ case.

We organize our paper as follows: in Section II we introduce basic notations and preliminaries; in section III we discuss the expectation value of the subsystem entropy of a random invariant tensor; in section IV we estimate the fluctuation; in section V a brief discussion is given.

\section{\label{sec:pre}Notations and preliminaries}

A multipartite quantum system of $n$-particles has a Hilbert space
$\mathcal{H}_n=\otimes_{i=1}^n V_{i}$, where all $V_{i}$ are identical with dimension $d=2j+1$.
An $n-$valent tensor is a vector $\ket{\psi_n}$ in $\mathcal{H}_n$. The spin angular momentum operators $\hat{J}_i^{a}$, acting on the $i$th particle read commutation relation
\begin{equation}
[\hat{J}^a_i,\hat{J}^b_j]=i\epsilon^{abc}\hbar\delta_{i,j} \hat{J}_i^c.
\end{equation}

Let the total angular momentum operator be $\hat{\mathbf{J}}=\sum_{i=1}^n \hat{\mathbf{J}}_i$. An $n-$valent tensor $\ket{\psi_n}$ is invariant if it satisfy
\begin{equation}
\mathbf{\hat{J}}\ket{\psi_n}=0.\label{invtensor}
\end{equation}

To add angular momentums, we use the standard Clebsch-Gordan
coefficients (CGCs) that are written as
\begin{equation}
C^{\, j_1\,\, j_2}_{m_1\, m_2\, J,\, M}=\langle j_1m_1;j_2m_2|J,M\rangle.
\label{3j-def}
\end{equation}
In order that the $3j$ symbol is nonzero, the spins $j_1,j_2,j_3$
have to satisfy the triangle inequality:
\begin{equation}\label{eq:triangular}
|j_1-j_2|\leq j_3\leq j_1+j_2.
\end{equation}
The CGCs can be chosen to be purely real by using proper relative phases. Moreover, the CGCs obey the following orthogonality relation
\begin{subequations}
\label{eq:orthogonal}
\begin{equation}
\sum_{m_1,m_2}C^{\, j_1\,\, j_2}_{m_1\, m_2\, J,\, M} C^{\, j_1\,\, j_2}_{m_1\, m_2\, J',\, M'}=\delta_{J,J'}\delta_{M,M'},
\end{equation}
\begin{equation}
\sum_{J,M}C^{\, j_1\,\, j_2}_{m_1\, m_2\, J,\, M} C^{\, j_1\,\, j_2}_{m'_1\, m'_2\, J,\, M}=\delta_{m_1,m'_1}\delta_{m_2,m'_2}.
\end{equation}
\end{subequations}

\begin{figure}[h]
\begin{center}
\subfigure[$\,\,$ trivalent tree]{
\label{figa} 
\includegraphics[width=5cm]{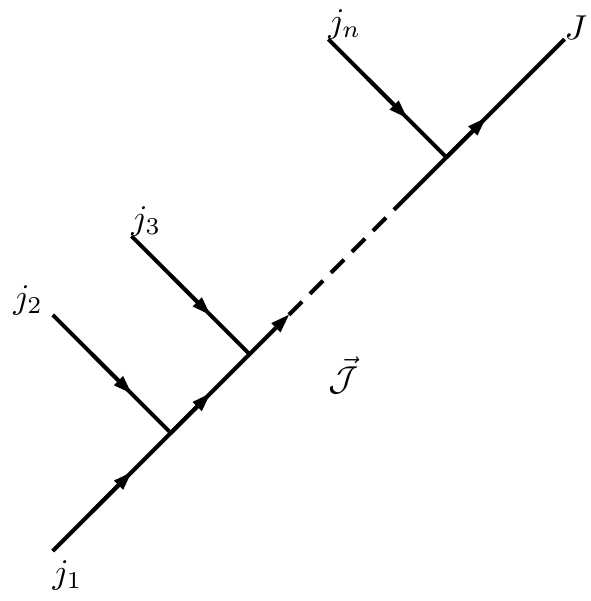}}
\hspace{2cm}
\subfigure[$\,\,$ node]{
\label{figb} 
\includegraphics[width=6cm]{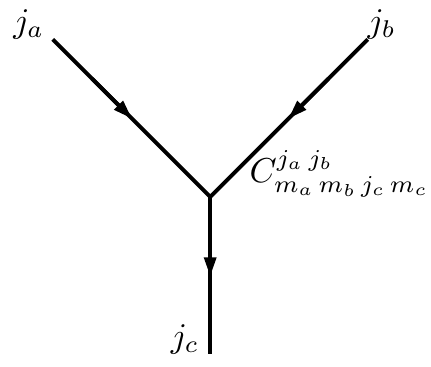}}
\label{tree} 
\caption{A tree made by contracting CGCs}
\end{center}
\end{figure}

$\mathcal{H}_n$ is invariant under global SU(2) action. Therefore, we could use tensor product of local basis (also known as uncoupling basis) to represent state $|\psi\rangle$ as,
\begin{equation}
|\psi\rangle=\sum_{\vec{m}}\psi_{\vec{m}}|\vec{m}\rangle,
\end{equation}
where multi-indices $\vec{m}$ stands for $\{m_1,m_2,\cdots,m_n\}$ and $\psi_{\vec{m}}$ is the corresponding coefficient; or equivalently, the coupling basis could be used
\begin{equation}
|\psi\rangle=\sum_{J,M,\vec{\mathcal{J}}}\psi_{J,M,\vec{\mathcal{J}}}|J,M,\vec{\mathcal{J}}\rangle,
\end{equation}
where $\{J, M\}$ is the total angular momentum and its $z$ components.
Notice that when $n>2$, ${J,M}$ are not enough to fully describe a state in $\mathcal{H}_n$, we have to add the multi-indices $\vec{\mathcal{J}}$ to indicate the specific coupling process, and hence uniquely determine the state. Of course these two representation can be related by unitary transformation
\begin{subequations}
\label{eq:unitary}
\begin{equation}
|\vec{m}\rangle=\sum_{J,M,\vec{\mathcal{J}}}\mathcal{T}^{J,M}_{\vec{m}}(\vec{\mathcal{J}})|J,M,\vec{\mathcal{J}}\rangle,
\end{equation}
\begin{equation}
|J,M,\vec{\mathcal{J}}\rangle=\sum_{\vec{m}}\mathcal{T}^{J,M}_{\vec{m}}(\vec{\mathcal{J}})^*|\vec{m}\rangle,
\end{equation}
\end{subequations}
where $\mathcal{T}^{J,M}_{\vec{m}}(\vec{\mathcal{J}})$ is a trivalent tree made by CGCs (see FIG.\ref{figa}), therefore every matrix element $\mathcal{T}^{J,M}_{\vec{m}}(\vec{\mathcal{J}})$ can take real numbers.The tree has $n+1$ external edges labeled by spins $\{\vec{j},J\}$. The internal edges are labeled by the set $\vec{\cj}$ of spins.

Since when $n>2$, the indices $\{J,M\}$ can not fully describe a state, there must exist degeneracy $D(j,n,J)$ for a label $\{J,M\}$. By definition,
\begin{equation}
D(j,n,J)=\sum_{\vec{\mathcal{J}}}1.
\end{equation}
It is straight forward from the triangular condition Eq.~(\ref{eq:triangular}) that $D(j,n,J)$ satisfies a recursive condition
\begin{eqnarray}\label{eq:D1}
\notag D(j,n,J)=
\sum_{J'=|J-j|}^{\min\left\{(J+j),(n-1)j\right\}}D(j,n-1,J'),\quad J\in \{0(\frac{1}{2}),1(\frac{3}{2}),\cdots,nj\}.\\
\end{eqnarray}
The number in bracket applies when $j$ is a half integer and $n$ is an odd number.
Notice that when $n$ is an odd number, $j$ can not be a half-integer.
The initial condition of recursion is that no degeneracy exsits
\begin{equation}\label{eq:D2}
D(j,2,J)=1,\quad J\in \{0,1,\cdots,2j\}.
\end{equation}

Combine Eq.~(\ref{eq:D1}) and (\ref{eq:D2}) together,it is easy to obtain two conclusion
\begin{itemize}
\item $D(j,n,J)$ is a polynomial of $j$ and $J$, and its highest power is $n-2$:
\begin{eqnarray}
D(j,n,J)=\sum_{k_1+k_2=0}^{n-2}c_{k_1,k_2}(n)j^{k_1}J^{k_2},
\end{eqnarray}
\item $D(j,n,J)$ is a step function that
\begin{eqnarray}
D(j,n,J)=\left\{
\begin{tabular}{ll}
$f_1(J,j,n)$ & $J\in\{0,1,\cdots,j\}$,\\
$f_2(J,j,n)$ & $J\in\{j+1,j+2,\cdots,2j\}$,\\
$\cdots$ &\\
$f_n(J,j,n)$ & $J\in\{(n-1)j+1,(n-1)j+2,\cdots,nj\}$,\\
\end{tabular}
\right.
\end{eqnarray}
here and hereafter we ignored the situation under which $j$ is half integer, because this will not affect our final result when discussing the asymptotical situation.
\end{itemize}

An $n$-valent tensor $\ket{\psi_n}$ is perfect, if
for any bipartition $A$ and $\bar{A}$, after tracing out the subsystem $A$, which is consisted of more than half(or equal) particles,
the entropy of the reduced density matrix $\rho_{\bar{A}}$ is maximal
\begin{equation}
\rho_{\bar{A}}=\text{Tr}_{A}\ket{\psi_n}\bra{\psi_n}=d^{-|\bar{A}|}\mathbb{I}_{d^|\bar{A}|},
\end{equation}
where $|\bar{A}|$ is the cardinality of $\bar{A}$.

The von Neumann entropy is defined by
\be
S=\text{Tr} \rho \ln\rho,
\ee
so tt is clear that the reduced density matrix $\rho_A$ of a perfect tensor processed entropy $n_A\ln(2j+1)$, which is the maximum entropy the subsystem can have, and we simply denote this as $S_{\max}(A)$. For arbitrary tensor which is not perfect, the entropy can not reach $S_{max}$. We simply define the average information of $A$ as the entropy deficit from $S_{\max}(A)$
\be
I(A)\equiv S_{\max}(A)-\overline{S(A)}
\ee
Our goal is to study the behavior of $I_{inv}(A)$, the entropy deficit of the reduced density matrix of a random tensor $\ket{\psi_n}$ in invariant subspace $\mathcal{H}_{n}^{(inv)}$ from the maximum could be, as $d=2j+1$, the dimension of local system approaches to infinity.
\section{\label{sec:mean}Second Renyi Entropy}

Given an invariant tensor $|I\rangle$, which lies in the invariant subspace $\mathcal{H}_{n}^{(inv)}=\mathrm{Inv}_{SU(2)}\mathcal{H}_n$, we define the density matrix $\rho=|I\rangle\langle I|$. We consider an arbitrary bipartition into $2$ sets of tensor legs. We consider the reduced density matrix $\rho_{A}=\trace_{\bar{A}}\rho$ by tracing out the degrees of freedom of set $A$. Without loss of generality,  we take $\bar{A}=\{n_A+1,n_A+2,\cdots,n\}$, which contains more than or equal to half of all particles, hence the remaining part $n_A\leq \lfloor\frac{n}{2}\rfloor$
The second Renyi entropy $S_2(A)$ of $\rho_{A}$ is given by
\begin{eqnarray}
e^{-S_2(A)}=\frac{\trace \rho_{A}^2}{(\trace \rho_{A})^2}.\label{S2}
\end{eqnarray}
By swapping trick, the numerator could be rewritten as
\begin{eqnarray}
Z_1\equiv \trace \rho_{A}^2=\trace\left[\left(\rho\otimes\rho\right)\mathcal{F}_{A}\right],
\end{eqnarray}
where the last trace is over the space $(\otimes_{i=1}^n V_{i})^{\otimes 2}$. $\mathcal{F}_{A}$ is a swap operator defined by
\begin{eqnarray}\notag
&&\mathcal{F}_{A}\bigg (|\vec{m}_A,\vec{m}_{\bar{A}}\rangle\otimes|\vec{m}'_A,\vec{m}'_{\bar{A}}\rangle\bigg)
=|\vec{m}'_A,\vec{m}_{\bar{A}}\rangle\otimes|\vec{m}_A,\vec{m}'_{\bar{A}}\rangle.\\
&&
\end{eqnarray}
The denominator of Eq.(\ref{S2}) can be written similarly as
\begin{eqnarray}
Z_0\equiv(\trace \rho_{A})^2=\trace\lt[\rho\otimes\rho\rt].
\end{eqnarray}

We random sample the invariant tensors $|I\rangle$ in the invariant subspace $\mathcal{H}_{n}^{(inv)}$, and consider the average
\begin{eqnarray}
\overline{Z_1}=\trace\left[\overline{\left(\rho\otimes\rho\right)}\mathcal{F}_{A}\right].
\end{eqnarray}
By Schur's Lemma \cite{church,Qi1}
\be\label{rho2}\notag
&&\overline{\rho\otimes\rho}\\
\notag &&\quad=\int\rmd U\, (U\otimes U)|0\rangle\langle 0|\otimes |0\rangle\langle 0|(U^\dagger\otimes U^\dagger)\\
&&\quad=\frac{1}{\dim(\mathcal{H}_{n}^{(inv)})^2+\dim(\mathcal{H}_{n}^{(inv)})}\lt(\ci+\cf\rt),
\ee
where $|0\rangle$ is an arbitrary reference state in $\mathcal{H}_{n}^{(inv)}$. The average is over all unitary operators $U$ on $\mathcal{H}_{n}^{(inv)}$. $\ci$ is the identity operator on $\mathcal{H}_{n}^{(inv)}\otimes\mathcal{H}_{n}^{(inv)}$, and $\cf$ is the swap operator
\be\notag
\ci |I\rangle\otimes |I'\rangle&=&|I\rangle\otimes |I'\rangle,\\
\cf |I\rangle\otimes |I'\rangle&=&|I'\rangle\otimes |I\rangle.
\ee

$\overline{Z_1}$ is computed as follows
\be
&&\lt[\dim(\mathcal{H}_{n}^{(inv)})^2+\dim(\mathcal{H}_{n}^{(inv)})\rt]\overline{Z_1}\nonumber\\
&&\quad=\sum_{\vec{m}_A,\vec{m}_{\bar{A}}}\sum_{\vec{m}'_A,\vec{m}'_{\bar{A}}}
\langle\vec{m}_A,\vec{m}_{\bar{A}}|\otimes\langle\vec{m}'_A,\vec{m}'_{\bar{A}}|\lt(\ci+\cf\rt)\cf_{A}
|\vec{m}_A,\vec{m}_{\bar{A}}\rangle\otimes|\vec{m}'_A,\vec{m}'_{\bar{A}}\rangle
\nonumber\\
&&\quad=\sum_{\vec{m}_A,\vec{m}_{\bar{A}}}\sum_{\vec{m}'_A,\vec{m}'_{\bar{A}}}
\langle\vec{m}_A,\vec{m}_{\bar{A}}|\otimes\langle\vec{m}'_A,\vec{m}'_{\bar{A}}|\lt(\ci+\cf\rt)|
\vec{m}'_A,\vec{m}_{\bar{A}}\rangle\otimes|\vec{m}_A,\vec{m}'_{\bar{A}}\rangle.
\ee

$\ci$ and $\cf$ act on the invariant tensors in $\mathcal{H}_{n}^{(inv)}\otimes\mathcal{H}_{n}^{(inv)}$. So when they acting on the right, they gives
\be
&&\lt(\ci+\cf\rt)|\vec{m}'_A,\vec{m}_{\bar{A}}\rangle\otimes|\vec{m}_A,\vec{m}'_{\bar{A}}\rangle\nonumber\\
&&\quad=\lt(\ci+\cf\rt) P_{inv}\otimes P_{inv}|\vec{m}'_A,\vec{m}_{\bar{A}}\rangle\otimes|\vec{m}_A,\vec{m}'_{\bar{A}}\rangle\nonumber\\
&&\quad=\sum_{I,I'}
|I\rangle I_{\vec{m}_{A}',\vec{m}_{\bar{A}}}
\otimes
|I'\rangle I'_{\vec{m}_{A},\vec{m}_{\bar{A}}'}+\sum_{I,I'}
|I\rangle I_{\vec{m}_{A},\vec{m}'_{\bar{A}}}
\otimes
|I'\rangle I'_{\vec{m}'_{A},\vec{m}_{\bar{A}}}
\ee
where we have used $I$ to label an orthonormal basis in $\mathcal{H}_{n}^{(inv)}$. $P_{inv}=\sum_{I}|I\rangle\langle I|$ is the projector onto the invariant subspace $\mathcal{H}_{n}^{(inv)}$. $I_{\vec{m}_{A},\vec{m}_{\bar{A}}}=\langle I|\vec{m}_{A},\vec{m}_{\bar{A}}\rangle$ is the invariant tensor component.

$\overline{Z_1}$ is thus expressed as
\be
&&\lt[\dim(\mathcal{H}_{n}^{(inv)})^2+\dim(\mathcal{H}_{n}^{(inv)})\rt]\overline{Z_1}\nonumber\\
&&\quad=\sum_{\vec{m}_A,\vec{m}_{\bar{A}}}\sum_{\vec{m}'_A,\vec{m}'_{\bar{A}}}
\left(\sum_{I,I'}
(I_{\vec{m}_{A},\vec{m}_{\bar{A}}})^*
I_{\vec{m}'_{A},\vec{m}_{\bar{A}}}
(I'_{\vec{m}'_{A},\vec{m}'_{\bar{A}}})^*
I_{\vec{m}_{A},\vec{m}'_{\bar{A}}}+\sum_{I,I'}
(I_{\vec{m}_{A},\vec{m}_{\bar{A}}})^*
I_{\vec{m}_{A},\vec{m}'_{\bar{A}}}
(I'_{\vec{m}'_{A},\vec{m}'_{\bar{A}}})^*
I_{\vec{m}'_{A},\vec{m}_{\bar{A}}}
\right)\nonumber\\ \label{avgZ1}
\ee
We choose the recoupling scheme that the orthonormal basis $|I\rangle$ can be written as
\be
I_{\vec{m}_{A},\vec{m}_{\bar{A}}}=
\sum_{M}\frac{(-1)^{J-M}}{\sqrt{2J+1}}
\ct^{J,M}_{\vec{m}_{A}}(\cj_A)\,\ct^{J,-M}_{\vec{m}_{\bar{A}}}(\cj_{\bar{A}}).
\ee
The internal edges are labeled by the set of spins $\cj_A$. The index $I$ is equivalent to the set of spin labels $(\cj_L,J,\cj_R)$.

Because of the orthogonality condition of CGCs, we have
\be
&&\sum_{\vec{m}_{A},\vec{m}_{\bar{A}}}
\ct^{J,M}_{\vec{m}_{A}}(\cj_A)^*\,
\ct^{J',M'}_{\vec{m}_{A}}(\cj_A')=\delta_{\cj_A,\cj_A'}\delta_{J,J'}\delta_{M,M'}.\notag\\
&&
\ee
which proves the orthonormality
\be
\sum_{\vec{m}_A,\vec{m}_{\bar{A}}}(I_{\vec{m}_{A},\vec{m}_{\bar{A}}})^*
I'_{\vec{m}_{A},\vec{m}_{\bar{A}}}=\delta_{I,I'}
\ee

Inserting into $\overline{Z_1}$, we find the first term in Eq.(\ref{avgZ1}) gives

\be
&&\sum_{\vec{m}_A,\vec{m}_{\bar{A}}}\sum_{\vec{m}'_A,\vec{m}'_{\bar{A}}}
\sum_{I,I'}
(I_{\vec{m}_{A},\vec{m}_{\bar{A}}})^*
I_{\vec{m}'_{A},\vec{m}_{\bar{A}}}
(I'_{\vec{m}'_{A},\vec{m}'_{\bar{A}}})^*
I'_{\vec{m}_{A},\vec{m}'_{\bar{A}}}\nonumber\\
&&\quad=\sum_{I,I'}\sum_{\vec{m},\vec{m}'}\sum_{M,\tilde{M}}\frac{(-1)^{2J-M-\tilde{M}}}{{2J+1}}
\sum_{N,\tilde{N}}\frac{(-1)^{2J'-N-\tilde{N}}}{{2J'+1}}\notag\\
&&\quad\quad\times\ct^{J,M}_{\vec{m}_{A}}(\cj_A)^*\,
\ct^{J,-M}_{\vec{m}_{\bar{A}}}(\cj_{\bar{A}})^*
\ct^{J,\tilde{M}}_{\vec{m}'_{A}}(\cj_A)\,
\ct^{J,-\tilde{M}}_{\vec{m}_{\bar{A}}}(\cj_{\bar{A}})\nonumber\\
&&\quad\quad\times
\ct^{J',N}_{\vec{m}_{A}'}(\cj'_A)^*\,
\ct^{J',-N}_{\vec{m}'_{\bar{A}}}(\cj'_{\bar{A}})^*
\ct^{J',\tilde{N}}_{\vec{m}_{A},\tilde{N}}(\cj'_A)\,
\ct^{J',-\tilde{N}}_{\vec{m}'_{\bar{A}}}(\cj'_{\bar{A}})\nonumber\\
&&\quad=\sum_{J,J'}\sum_{\cj_{\bar{A}},\cj_{\bar{A}}'}\sum_{\cj_A,\cj_A'}\sum_{M,\tilde{M}}\sum_{N,\tilde{N}}
\frac{\delta_{M,\tilde{M}}\delta_{N,\tilde{N}}\delta_{J,J'}\delta_{M,\tilde{N}}\delta_{\tilde{M},{N}}\delta_{\cj_A,\cj_A'}}{(2J+1)^2}\nonumber\\
&&\quad=\sum_{\cj_{\bar{A}},\cj_{\bar{A}}'}\sum_{\cj_A}\sum_{J}(2J+1)^{-1}.
\ee

Note that the sums over $\cj_{\bar{A}},\cj_A,J$ are not independent, but constrained by triangle inequalities at each node in the tree.

The second term in Eq.(\ref{avgZ1}) gives the similar result by $\cj_A\leftrightarrow \cj_{\bar{A}}$
\be
\sum_{\cj_A,\cj_A'}\sum_{\cj_{\bar{A}}}\sum_{J}(2J+1)^{-1}
\ee

Therefore
\begin{equation}
\lt[\dim(\mathcal{H}_{n}^{(inv)})^2+\dim(\mathcal{H}_{n}^{(inv)})\rt]\overline{Z_1}
=\sum_{\cj_{\bar{A}},\cj_{\bar{A}}'}\sum_{\cj_A}\sum_{J}(2J+1)^{-1}
+\sum_{\cj_A,\cj_A'}\sum_{\cj_{\bar{A}}}\sum_{J}(2J+1)^{-1}.
\end{equation}
where
\be
\dim(\mathcal{H}_{n}^{(inv)})=\sum_{\cj_R}\sum_{\cj_L}\sum_{J}1
\ee
Again the sums are constrained by triangle inequalities.


The average of $Z_0$ can be computed in a similar way, by removing the swap $\cf_{A}$
\be
&&\lt(\dim(\mathcal{H}_{n}^{(inv)})^2+\dim(\mathcal{H}_{n}^{(inv)})\rt)\overline{Z_0}\notag\\
&&\quad=
\sum_{\vec{m},\vec{m}'}\langle\vec{m}|\otimes\langle\vec{m}'|(\ci+\cf)|\vec{m}\rangle\otimes |\vec{m}'\rangle\nonumber\\
&&\quad=\sum_{\vec{m},\vec{m}'}
\Bigg(\sum_{I,I'}(I_{\vec{m}})^* I_{\vec{m}}(I'_{\vec{m}'})^* I'_{\vec{m}'}+
\sum_{I,I'} (I_{\vec{m}})^* I_{\vec{m}'} (I'_{\vec{m}'})^* I'_{\vec{m}}\Bigg)\nonumber\\
&&\quad=1.
\ee
In fact, if we require that the density matrix to be normalized, we can obtain $\overline{Z_0}=1$ immediately.

Recall that $D(j,n,J)$ is the dimension of the space constituted of $n$-valent invariant tensors with each subsystem of spin $j$. Then
\be
\sum_{\cj_{\bar{A}},\cj'_{\bar{A}}}\sum_{\cj_A}\sum_{J}(2J+1)^{-1}&=&
\sum_{J}\frac{D(j,n_{\bar{A}},J)^2D(j,n_A,J)}{2J+1}\nonumber\\
\sum_{\cj_A,\cj_A'}\sum_{\cj_{\bar{A}}}\sum_{J}(2J+1)^{-1}&=&
\sum_{J}\frac{D(j,n_{\bar{A}},J)D(j,n_A,J)^2}{2J+1}\nonumber\\
\dim(\mathcal{H}_{n}^{(inv)})&=&\sum_{J}{D(j,n_{\bar{A}},J)D(j,n_A,J)}.\notag\\
\label{dim_H}
\ee

Hence, $\dim(\mathcal{H}_{n}^{(inv)})$ is a polynomial of $j$, whose highest power is $n-3$:
\be
\dim(\mathcal{H}_{n}^{(inv)})\sim j^{n-3},
\ee
which can be observed directly from Eq.(\ref{dim_H}).

Setting $J=0$, $D(j,n,J)$ will reduce to $\dim(\mathcal{H}_{n}^{(inv)})$ and become a polynomial with highest power $n-3$. However, $D(j,n,J)$ itself is a polynomial of highest power $n-2$, therefore, we know that, $D(j,n,J)$ does not contain $c_{n-2,0}(n)j^{n-2}$ in its highest order terms when $J\in \{0,1,\cdots,j\}$.

Setting $j\rightarrow \infty$,
\be
\overline{S_2(A)}
&=&\lim_{j\rightarrow \infty}2\ln\lt(\sum_{J}{D(j,n_{\bar{A}},J)D(j,n_A,J)}\rt)\notag\\
&&+\lim_{j\rightarrow \infty}\ln\lt(1+\frac{1}{\sum_{J}{D(j,n_{\bar{A}},J)D(j,n_A,J)}}\rt)\notag\\
& &-\lim_{j\rightarrow \infty}\ln\lt(\lambda\sum_{J=0}^{nj}\frac{D(j,n_A,J)D(j,n_{\bar{A}},J)^2}{2J+1}\rt).
\ee

Because both $D(j,n_A,J)$ and $D(j,n_{\bar{A}},J)$ contain only finite terms, when we consider the asymptotic behaviour of $\overline{S_2(A)}$, only the leading term has contribution. (The rest ones appearing in the summation would contribute $\ln(1+ \textrm{others}/\textrm{leading})$ and hence vanish.)

The leading term of $\sum_{J}{D(j,n_{\bar{A}},J)D(j,n_A,J)}$ is of form
\be
C_1j^{2*(n-3)},
\ee
where $C_1$ is a constant stemming from summation, which does not depend on $j$.

The power of $J$ in the leading term of
\[\sum_{J=0}^{nj}\frac{D(j,n_A,J)D(j,n_{\bar{A}},J)^2}{2J+1}\]
is at least
\[
\min\{n_A-2,1\}+2*\min\{n_{\bar{A}}-2,1\}-1,
\]
when $J\in\{0,1,\cdots,j\}$.

When $N \geq 5$, it is positive and thus we will obtain a higher order polynomial after the summation over $J$. However the power in numerator could be $0$ when $J>j$, therefore after summation, harmonic number would arise. Though such harmonic numbers depend on $j$ and are divergent, they cancel each other and the result will converge to a finite number when $j$ goes to infinity. The leading term of \[\sum_{J=0}^{nj}\frac{D(j,n_A,J)D(j,n_{\bar{A}},J)^2}{2J+1}\]
should be of
\be
\lambda_n C_2 j^{n_A-2+2*(n_{\bar{A}}-2)},
\ee
where $C_2$ is a coefficient stemming from summation, which will converge to a finite number as $j\rightarrow\infty$, and $\lambda_n$ is the multiplicity, $1$ for odd $n$ and $2$ for the even case.

Therefore,
\be
\overline{S_2(A)}&\rightarrow& n_A\ln j+\ln\frac{C_1}{\lambda_n C_2}.
\ee
Because the von Neumann entropy is lower bounded by the second Renyi, $S\geq S_2$ and $S_{\max}=n_A\ln (2j+1)$ is the entropy of $n-$valent perfect tensor. The average information of subsystem is
\be
I_{inv}(A)=S_{\max}-\overline{S(A)}\leq n_A\ln 2 -\ln\frac{C_1}{\lambda_n C_2},
\ee
here $I_{inv}(A)$ is only a finite number which only depends on $n$.

If the average runs over the whole Hilbert space $\mathcal{H}_n$, the average information of subsystem is
\be
I(A)=\frac{\dim(\otimes_{i=1}^{n_A} V_{i})}{2\dim(\otimes_{i=n_A+1}^n V_{i})}=\frac{1}{2}(2j+1)^{n_A-n_{\bar{A}}}.
\ee

The following table gives the numerical results when $n$ varies from $4$ to $14$ with $j=10^{100}$.\\
\begin{displaymath}
\begin{tabular}{c c c c l}
&    $n$ & $n_A$ & $I_{inv}(A)$ & $I(A)$\\
\hline
&   4   &   2   &   8.096     &   0.5\\
&	5	&	2	&	0.179     &   $\epsilon$\\
&	6	&	3	&	0.348     &   0.5\\
&	7	&	3	&	0.240     &   $\epsilon$\\
&	8	&	4	&	0.365     &   0.5\\
&	9	&	4	&	0.290     &   $\epsilon$\\
&	10	&	5	&	0.414     &   0.5\\
&	11	&	5	&	0.350     &   $\epsilon$\\
&	12	&	6	&	0.493     &   0.5\\
&	13	&	6	&	0.415     &   $\epsilon$\\
&	14	&	7	&	0.544     &   0.5\\
\end{tabular}
\end{displaymath}
where $\epsilon=5\times 10^{-101}$.

It is surprisingly to notice that, when $n$ is a small even number ($n >4$) and $n_A=\frac{n}{2}$, the average information of $A$ over invariant subspace is even smaller than the average over the whole Hilbert space. We take this as coincidence but no more profound reason behind.

\section{Bound on Fluctuations}

In this section, we estimate the bound on fluctuation of Renyi entropy $S_2(A)$ around the average $\overline{S_2(A)}$, under the asymptotical limit that $j\to\infty$. Using the bound, we also show that $S_2(A)$ concentrates at $\overline{S_2(A)}$ with a high probability, which is close to 1 as every $j\to\infty$. The idea of derivation is similar to \cite{Qi1}.

We consider the fluctuation:
\begin{equation}
\frac{\overline{(Z_{1,0}-\overline{Z_{1,0}})^2}}{\overline{Z_{1,0}}^2}=\frac{\overline{Z_{1,0}^2}}{\overline{Z_{1,0}}^2}-1.
\end{equation}

By Schur's Lemma
\be\label{rho4}\notag
&&\overline{\otimes^4\rho}=\frac{1}{\cc_4}\sum_{g\in S_4}g,
\ee
where
\[\cc_n=\frac{(\dim(\mathcal{H}_{n}^{(inv)})+n-1)!}{(\dim(\mathcal{H}_{n}^{(inv)})-1)!}.\]
Using swapping trick, we can rewrite
\be
\overline{Z_{1}^2}=\trace\lt[\lt(\overline{\otimes^{4}\rho}\rt)\cf_{A}\otimes\cf_{A}\rt],
\ee
where the two $\cf_{A}$ act on the subspace of the first and second $N$ particles respectively.
Then we can calculate ${\cc_{4}}\overline{Z_{1}^2}$ as
\begin{widetext}
\be
{\cc_{4}}\overline{Z_{1}^2}
&=&\sum_{\vec{m}_A^{(i)},\vec{m}_{\bar{A}}^{(i)}}
\bigotimes_{i=1}^{2}\langle\vec{m}_A^{(i)},\vec{m}_{\bar{A}}^{(i)}|\otimes\langle\vec{m}_A^{(2+i)},\vec{m}_{\bar{A}}^{(2+i)}|
\lt(\sum_{g\in S_{4}}g \cf_{A}\otimes\cf_{A}\rt)
\bigotimes_{i=1}^{2}|\vec{m}_A^{(i)},\vec{m}_{\bar{A}}^{(i)}\rangle\otimes|\vec{m}_A^{(2+i)},\vec{m}_{\bar{A}}^{(2+i)}\rangle\notag\\
&=&\sum_{g\in S_{4}}\sum_{\vec{m}_A^{(i)},\vec{m}_{\bar{A}}^{(i)}}
\bigotimes_{i=1}^{2}\langle\vec{m}_A^{(i)},\vec{m}_{\bar{A}}^{(i)}|\otimes\langle\vec{m}_A^{(i+2)},\vec{m}_{\bar{A}}^{(i+2)}|\notag\\
&&\times\lt(\bigotimes_{k=1}^{4}\sum_{I^{(k)}}|I^{(k)}\rangle\langle I^{(k)}|\rt)
\bigotimes_{i=1}^{2}|\vec{m}_A^{(g(i+1))},\vec{m}_{\bar{A}}^{(g(i))}\rangle\otimes |\vec{m}_A^{(g(i+3))},\vec{m}_{\bar{A}}^{(g(i+2))}\rangle\notag\\
&=&\sum_{g\in S_{4}}\prod_{i=1}^2\prod_{k=1}^2\sum_{\vec{m}_A^{(i)},\vec{m}_{\bar{A}}^{(i)}}\sum_{I^{(k)}}
\lt(I^{(k)}_{\vec{m}^{(i)}_{A},\vec{m}^{(i)}_{\bar{A}}}\rt)^*
\lt(I^{(k)}_{\vec{m}^{(i+2)}_{A},\vec{m}^{(i+2)}_{\bar{A}}}\rt)^*
I^{(k)}_{\vec{m}^{(g(i+1))}_{A},\vec{m}^{(g(i))}_{\bar{A}}}
I^{(k)}_{\vec{m}^{(g(i+3))}_{A},\vec{m}^{(g(i+2))}_{\bar{A}}}\notag\\
&=&\sum_{g\in S_{4}}\prod_{i=1}^2\prod_{k=1}^2\sum_{\vec{m}_A^{(i)},\vec{m}_{\bar{A}}^{(i)}}\sum_{I^{(k)}}
\sum_{M^{(k)}}\frac{(-1)^{J^{(k)}-M^{(k)}}}{\sqrt{2J^{(k)}+1}}
\lt(\ct^{J^{(k)},M^{(k)}}_{\vec{m}^{(i)}_{A}}(\cj_A)\rt)^*
\lt(\ct^{J^{(k)},-M^{(k)}}_{\vec{m}^{(i)}_{\bar{A}}}(\cj_{\bar{A}})\rt)^*\notag\\
&&\times\sum_{M^{(2+k)}}\frac{(-1)^{J^{(2+k)}-M^{(2+k)}}}{\sqrt{2J^{(2+k)}+1}}
\lt(\ct^{J^{(2+k)},M^{(2+k)}}_{\vec{m}^{(2+i)}_{A}}(\cj_A)\rt)^*
\lt(\ct^{J^{(2+k)},-M^{(2+k)}}_{\vec{m}^{(2+i)}_{\bar{A}}}(\cj_{\bar{A}})\rt)^*\notag\\
&&\times\sum_{\tilde{M}^{(k)}}\frac{(-1)^{J^{(k)}-\tilde{M}^{(k)}}}{\sqrt{2J^{(k)}+1}}
\lt(\ct^{J^{(k)},\tilde{M}^{(k)}}_{\vec{m}^{(g(i+1))}_{A}}(\cj_A)\rt)
\lt(\ct^{J^{(k)},-\tilde{M}^{(k)}}_{\vec{m}^{(g(i))}_{\bar{A}}}(\cj_{\bar{A}})\rt)\notag\\
&&\times\sum_{\tilde{M}^{(2+k)}}\frac{(-1)^{J^{(2+k)}-\tilde{M}^{(2+k)}}}{\sqrt{2J^{(2+k)}+1}}
\lt(\ct^{J^{(2+k)},\tilde{M}^{(2+k)}}_{\vec{m}^{(g(3+i))}_{A}}(\cj_A)\rt)
\lt(\ct^{J^{(2+k)},-\tilde{M}^{(2+k)}}_{\vec{m}^{(g(2+i))}_{\bar{A}}}(\cj_{\bar{A}})\rt)\notag\\
&=&\lt(\sum_{J}\frac{D(j,n_A,J)D(j,n_{\bar{A}},J)^2}{2J+1}
+\sum_{J}\frac{D(j,n_A,J)^2D(j,n_{\bar{A}},J)}{2J+1}\rt)^2\notag\\
&&+\sum_{g\not\in S_2\otimes S_2}\sum_{J}\frac{D(j,n_A,J)^{\chi(g\cdot \cf_{A}\otimes\cf_{A})}D(j,n_{\bar{A}},J)^{\chi(g)}}{(2J+1)^2},
\ee
\end{widetext}
where $\chi(g)$ is the number of disjoint circles of group element $g$.

Given that $\frac{(\cc_2)^2}{\cc_{4}}< 1$,
\be
\overline{\lt(\frac{Z_{1}}{\overline{Z_1}}-1\rt)^2}
=\frac{\overline{(Z_{1})^2}}{\overline{Z_{1}}^2}-1<f(j,n),
\ee
where
\begin{widetext}
\be
f(j,n)=\frac{\sum_{g\not\in S_2\otimes S_2}\sum_{J}(2J+1)^{-2}D(j,n_A,J)^{\chi(g\cdot \cf_{A}\otimes\cf_{A})}D(j,n_{\bar{A}},J)^{\chi(g)}}
{\lt(\sum_{J}\frac{D(j,n_A,J)D(j,n_{\bar{A}},J)^2}{2J+1}
+\sum_{J}\frac{D(j,n_A,J)^2D(j,n_{\bar{A}},J)}{2J+1}\rt)^2}\sim\frac{1}{j}
\ee
\end{widetext}

On the other hand,
\be
\overline{(Z_{0})^2}
&=&\frac{1}{\cc_{4}}\sum_{g\in S_{4}}\prod_{i=1}^{4}
\sum_{I^{(i)}}\delta^{I^{g(i)},I^{(i)}}\nonumber\\
&\simeq &\frac{1}{\cc_{4}}\lt[\dim(\mathcal{H}_{n}^{(inv)})^{4}+6\dim(\mathcal{H}_{n}^{(inv)})^{3}+\cdots\rt]
\ee
Therefore
\be
\overline{\lt(\frac{Z_{0}}{\overline{Z_0}}-1\rt)^2}<6\dim(\mathcal{H}_{n}^{(inv)})^{-1}\sim\frac{1}{j^{n-3}}.
\ee

By Markov's inequality,
\be
\mathrm{Prob}\lt(\lt|\frac{Z_{1}}{\overline{Z_1}}-1\rt|\geq\frac{\delta}{4}\rt)\leq \frac{\overline{\lt(\frac{Z_{1}}{\overline{Z_1}}-1\rt)^2}}{\lt(\frac{\delta}{4}\rt)^2}
\sim&\frac{1}{\delta^2}O(\frac{1}{j}) .\label{prob1}
\ee
and
\be
&&\mathrm{Prob}\lt(\lt|\frac{Z_{0}}{\overline{Z_0}}-1\rt|\geq\frac{\delta}{4}\rt)<\frac{96}{\delta^2}
\dim(\mathcal{H}_{n}^{(inv)})^{-1}
\sim \frac{1}{\delta^2}O(\frac{1}{j^{n-3}})\notag\\
&&.\label{prob2}
\ee

The bounds Eq.(\ref{prob1}) and (\ref{prob2}) imply that with the probability of at least $1-\frac{1}{\delta^2 j}$, we have $\lt|\frac{Z_{0,1}}{\overline{Z_{0,1}}}-1\rt|\leq\frac{\delta}{4}$. Then we have
\be
\lt|S_2(A)-\overline{S_2(A)}\rt|&=&\lt|\ln\frac{{Z_1}}{{Z_0}}-\ln\frac{\overline{Z_1}}{\overline{Z_0}}\rt|\notag\\
&=&\lt|\ln\frac{{Z_1}}{\overline{Z_1}}-\ln\frac{{Z_0}}{\overline{Z_0}}\rt|\notag\\
&\leq&\lt|\ln\frac{{Z_1}}{\overline{Z_1}}\rt|+\lt|\ln\frac{{Z_0}}{\overline{Z_0}}\rt|\nonumber\\
&\leq&\frac{\delta}{2}+\frac{\delta}{2}=\delta
\ee
where we have used that for $\delta\leq 2$, $|\ln(1\pm \delta/4)|\leq \delta/2$.

\section{Discussion}

In this work, we investigate the asymptotical behavior of the entanglement for the random tensor with arbitrary rank $n\geq5$ in the SU(2) invariant subspace when the local dimensions are large. The situation with lower rank random tensors has been studied in \cite{Li2016Invariant}. We show that in general the entanglement entropy of random invariant tensor is maximal asymptotically for large local dimensions. In other words, the random invariant tensor approximates an invariant perfect tensor. When $n\geq 5$, unlike the divegent subleading term in $n=4$ case, the entropy deficit in the entanglement entropy is finite, thus the random invariant tensor with $n\geq5$ is even closer to a perfect state. The average information is a finite number which grows with $n$. Under special situation when $n$ is a small even number and $n_A=\frac{n}{2}$, the average information is even smaller than the average value over the whole space. Indeed these $n-$valent tensors are highly entangled states.

The next natural questions would be what the maximum subsystem entanglement entropy within such invariant subspace can achieve, what such states look like and how about their measure comparing with the whole space. We leave these questions for future research.

As another future direction, the results in this work may be applied to the studies of random tensor networks \cite{Qi1,QiYY,Han:2017uco}, which are models to realize the holographic correspondence. The random invariant tensor has SU(2) invariance and can build random tensor networks with local SU(2) symmetry. The local symmetry of tensor network might be used to realize the bulk gauge symmetry in the holographic correspondence.

\section*{Acknowledgments}
We thank Markus Grassl and Jie Zhou for helpful discussions. Y.~L. acknowledges support from Chinese Ministry of Education under grants No.20173080024. M.~H. acknowledges support from the US National Science Foundation through grant PHY-1602867, and startup grant at Florida Atlantic University, USA.
B.~L. is supported by NSERC and CIFAR.

\end{document}